IEEE Computer Magazine - Software Engineering Column

# From Untestable to Testable: Metamorphic Testing in the Age of LLMs

Valerio Terragni, *University of Auckland*

*Abstract—This article discusses the challenges of testing software systems with increasingly integrated AI and LLM functionalities. LLMs are powerful but unreliable, and labeled ground truth for testing rarely scales. Metamorphic Testing solves this by turning relations among multiple test executions into executable test oracles.*

Large Language Models (LLMs) are eating the world. They are being embedded into everyday software at an incredible pace – not just as chatbots anymore, but as part of real software products and real decisions. On the surface, that is exciting – LLM-powered software promises smarter tools, new features, and more automation. It is also *dangerous*. LLMs can hallucinate, contradict themselves, and deliver wrong answers with complete confidence. This leaves many researchers and CTOs asking: How do we test this new generation of LLM-powered software?

Testing any software system requires **test oracles** – something that tells you whether a test passed or failed. The most natural oracle for testing any software system is ground-truth labeled data. In sentiment analysis, for instance, you can manually label each input as positive or negative; any disagreement between the model output and the label is a test failure. The same applies for classical software – you know what a function should return, and you check whether it does. In principle, it's clean and simple. In practice, the scale requirements make this approach quickly collapse. LLM behavior is extraordinarily sensitive to input phrasing and failures can be sparse and subtle, so gaining real confidence often requires testing across millions of inputs. Hiring enough annotators to label all this is not just expensive, but often simply not feasible.

And the problem runs deeper still. LLMs have changed how we build AI systems – instead of laboriously assembling large labeled datasets to train models from scratch, we can now just use a pre-trained model instead. Add to that a few examples (few-shot learning) or a retrieval-augmented pipeline (RAG) over existing documents, and you have no need for extensive data. Convenient for development; but if we had no labeled data for training, we (usually) have no labeled data for testing.

Four properties of LLMs conspire to make this worse: open-ended outputs, multiple valid responses per input, high context-sensitivity, and non-determinism. Together, it means human labelers may not even agree on what "correct" looks like. Classic NLP benchmarks exist, but those rarely match the deployment context – in practice, inputs may be too domain-specific or simply not covered. New AI-driven applications are emerging faster than any benchmark effort can follow. We cannot wait for labeled data to appear for every task LLMs are being asked to perform. So how can we automatically test LLM-powered systems at scale, without ground-truth labels?

## Metamorphic Testing: One Oracle for Many Tests

The answer I find most compelling was actually introduced decades before LLMs existed. **Metamorphic Testing (MT)** [1] was proposed in 1998 by T.Y. Chen, S.C. Cheung (my PhD advisor 2012–2017), and S.M. Yiu as a way to address the oracle problem. Here, we shift away from ground-truth test oracles and toward *relation oracles*. The core idea is elegant: for each input, instead of asking "What should be the correct output?", we ask "How should the outputs change (or stay the same) when we make a systematic change to the input?" [2].

The intuition is something most experienced engineers will recognize immediately: You may not know the right answer for a specific input, but you often do know how the result should look like as inputs change. If I negate a number, squaring it should give the same result: $x_1 = -x_2 \Rightarrow x_1^2 = x_2^2$. That's a trivial example, but the principle scales [2]. These expected relationships are called **metamorphic relations** (MRs): If a certain input relation $R_i$ holds, a corresponding output relation $R_o$ must hold; formally, $R_i \Rightarrow R_o$.

This is the key economic advantage (and the connection to the "million labels" problem from the previous section). Defining a metamorphic relation is a one-time effort, and once defined, it can be applied broadly. MT starts by generating a *source* test input, then applies the MR input transformation to create a *follow-up* input (such that the source and follow-up input satisfies the input relation), executes the system on both inputs, and reports a failure if the expected output relation is violated. In other words, *one MR can act as an oracle for an arbitrary number of automatically generated source and follow-up tests*, potentially scaling to millions of test cases without requiring a single ground-truth label.

This is not an academic curiosity. Currently, Google Scholar has indexed 5,450+ entries mentioning "metamorphic testing", and MT has been deployed in practice at large organizations including Meta [4], NASA [5], and Google [6]. The question is: Can it be used to test LLM-powered software? Can it help resolve the crisis of labeled data and the need for cheap ways to automatically test such systems?

## Metamorphic Testing for LLMs

Despite its potential, MT for LLMs remains surprisingly understudied. Until recently, only a handful of papers had explored the intersection, and none had done so comprehensively. To address this, my group conducted what is, to date, the most exhaustive study of MT for LLMs [7]. As MT is already well-established in NLP, we systematically reviewed 1,024 papers and extracted 191 MRs for NLP spanning 24 tasks (e.g., sentiment analysis, question answering, relation extraction)[1]. Some of the 191 MRs are simple and general: *if you paraphrase the input, the output should stay the same*. Others are more semantically subtle: *if you swap two entities that have an asymmetrical relation, the predicted relation should flip accordingly*.

We implemented 36 of these MRs in a framework called LLMORPH [9] and ran ~560,000 metamorphic tests across three popular LLMs (GPT-4, LLAMA 3, and HERMES-2). The results were a mixed bag of genuine opportunities and cautionary truths.

**On the positive side**, MT successfully exposed faulty behaviors across all three models, with an average failure rate of 18% across the 36 implemented MRs, and up to 80% for individual relations. We also found that several task-independent MRs proved effective across multiple tasks.

**On the negative side**, MT in NLP inherits the ambiguity of language, making false positives hard to avoid. Transformations may shift meaning unintentionally (a misspelling that changes meaning, a paraphrase that shifts semantics), outputs can differ yet both be acceptable, and deciding whether two outputs are "the same" often requires semantic comparison; thus, our BERT-based similarity score effectively becomes part of the oracle. Our manual analysis of 937 violations found a true positive rate of around 62%. Not perfect, but comparable to traditional MT for NLP – LLMs do not make this problem worse, even if they do not make it better.

A subtler challenge is what to do once you find a failure. In classical software testing, a failing test points to a specific code location; then, you debug it, fix it, and verify the fix. With LLM-powered systems, the causal chain is not that easy. A metamorphic violation might be addressed on the software side (adding guardrails, refining the prompt, filtering outputs), but the root cause may lie at the model level, embedded in billions of parameters trained on data you never saw. In that case, the "fix" may require targeted fine-tuning and alignment adjustments.

## A Roadmap and a Call to Action

We have only scratched the surface of what MT can offer for LLM-powered software. One practical takeaway for teams deploying LLMs today is thus: a single well-chosen MR is a one-time investment that can generate millions of test cases. This is almost certainly cheaper than labeling even a fraction of them. These MR-based test suites can then serve as lightweight regression signals – run them whenever the model changes, the prompt changes, or the deployment context shifts, and watch for drift in failure rate (outside some unavoidable false positives).

But the deeper opportunity lies beyond natural language. LLMs are now among the most prolific generators of code [8], and code comes with properties that language does not – it compiles or it doesn't; it passes tests or it fails them; issues can be automatically flagged by static analyzers. These properties can enable better output comparisons with fewer false positives. Further out, new MRs can be explored in the context of the ever-more-popular agentic AI systems (e.g., changing the order of agent invocations should not alter the final answer under well-defined constraints).

My **call to action** is addressed to two audiences. **Practitioners** do not need to wait for the research community to mature this field. Starting with five to ten MRs specific to your system, integrating MT into existing CI pipelines, and tracking failure-rate trends over time can greatly increase the quality of your systems. **Researchers**, meanwhile, have rich open problems ahead: better input transformations that preserve semantics more reliably, better output comparators that handle the ambiguity of natural language, automatic MR discovery, and new MRs for agentic systems.

---

[1] https://mt4nlp.github.io/

As LLMs creep into our software and start powering real features and decisions, the question of how to validate their behavior at scale becomes urgent. Metamorphic testing will not solve this problem on its own – but it is one of the sharpest tools we have right now, and we are only beginning to learn how to use it.

## REFERENCES

1. T. Y. Chen, S. C. Cheung, and S. M. Yiu, "Metamorphic testing: A new approach for generating next test cases," Tech. Rep. HKUST-CS98-01, Dept. of Computer Science, The Hong Kong University of Science and Technology, 1998 (Technical Report)
2. T. Y. Chen, F.-C. Kuo, H. Liu, P.-L. Poon, D. Towey, T. H. Tse, and Z. Q. Zhou, "Metamorphic testing: A review of challenges and opportunities," *ACM Computing Surveys*, vol. 51, no. 1, pp. 4:1–4:27, 2018 (Journal)
3. X. Hou, Y. Zhao, Y. Liu, Z. Yang, K. Wang, L. Li, X. Luo, D. Lo, J. Grundy, and H. Wang. "Large language models for software engineering: A systematic literature review," ACM Trans. on Soft. Eng. and Meth. vol. 33, no. 8, 2024 (Journal)
4. J. Ahlgren, M. E. Berezin, K. Bojarczuk, E. Dulskyte, I. Dvortsova, J. George, N. Gucevska, M. Harman, M. Lomeli, E. Meijer, S. Sapora, and J. Spahr-Summers, "Testing Web Enabled Simulation at Scale Using Metamorphic Testing," in *Proc. IEEE/ACM 43rd Int. Conf. on Software Engineering: Software Engineering in Practice (ICSE-SEIP)*, pp. 140–149, 2021, doi: 10.1109/ICSE-SEIP52600.2021.00023 (Conference)
5. M. Lindvall, D. Ganesan, R. Ardal, and R. E. Wiegand, "Metamorphic Model-Based Testing Applied on NASA DAT: An Experience Report," in *Proc. IEEE/ACM 37th Int. Conf. on Software Engineering (ICSE)*, pp. 129–138, 2015 (Conference)
6. A. F. Donaldson, H. Evrard, and P. Thomson. "Putting randomized compiler testing into production (experience report)," in *Proc. 34th Eur. Conf. Object-Oriented Programming (ECOOP)*, Berlin, Germany (virtual), Nov. 15–17, 2020, LIPIcs, vol. 166, Art. no. 22, pp. 22:1–22:29. (Conference)
7. S. Cho, S. Ruberto, and V. Terragni, "Metamorphic Testing of Large Language Models for Natural Language Processing," in *Proc. IEEE Int. Conf. on Software Maintenance and Evolution (ICSME)*, 2025, doi: 10.1109/ICSME64153.2025.00025 (Conference)
8. V. Terragni, A. Vella, P. Roop, and K. Blincoe, "The Future of AI-Driven Software Engineering," *ACM Trans. Softw. Eng. Methodol.*, vol. 34, no. 5, Art. 120, May 2025, doi: 10.1145/3715003 (Journal)
9. S. Cho, S. Ruberto, and V. Terragni, "LLMORPH: Automated Metamorphic Testing of Large Language Models" in *Proc. Int. Conf. on Automated Software Engineering (ASE)*, 2025, (Conference).

**Valerio Terragni** is a Senior Lecturer (equivalent to an Associate Professor in U.S., Asian, and European systems) at The University of Auckland, Auckland 1010, New Zealand. Contact him at v.terragni@auckland.ac.nz